\documentclass{article}
\usepackage{spconf,amsmath,graphicx,hyperref}
\usepackage{cite,xcolor,amsmath,tikz, etoolbox}
\usepackage{amsmath,amssymb,amsfonts,algorithmic,graphicx,textcomp,xcolor,booktabs,tabularx,makecell,multirow}
\def\BibTeX{{\rm B\kern-.05em{\sc i\kern-.025em b}\kern-.08em
    T\kern-.1667em\lower.7ex\hbox{E}\kern-.125emX}}

\usepackage{pgfplots}
\pgfplotsset{compat=1.18} 
\DeclareMathOperator*{\argmax}{\arg\!\max}
\usetikzlibrary{positioning,shadows,arrows,arrows.meta,shapes,calc,chains}

\title{LEARNING NEW WORDS FROM UNLABELED TEST DATA IN AUTOMATIC SPEECH RECOGNITION}
\name{Mengqi Wang$^{\star}$ \qquad Mark A. Hasegawa-Johnson$^{\star}$ \qquad Haolong Zheng$^{\star}$ \qquad Chang D. Yoo$^{\dagger}$}
\address{$^{\star}$ University of Illinois at Urbana-Champaign, USA \\
         $^{\dagger}$ Korea Advanced Institute of Science and Technology, South Korea \\
         \texttt{\small{\{jhasegaw, mengqiw3, haolong2\}@illinois.edu, cd\_yoo@kaist.ac.kr}}
        }
\begin{document}
\ninept
\maketitle
\begin{abstract}
New words are invented every day. A human listener can learn a new word by hearing it clearly once and inferring its usage from sentence context. This paper proposes granting ASR a similar ability to learn the contextual representations and spellings of new words from unlabeled test data at test time. A frozen CTC acoustic model provides spellings, a frozen language model provides contextual evidence for out-of-vocabulary (OOV) word detection, and an adaptation module expands the vocabulary by learning the lexical token representations with distributions over CTC-generated candidates. The spelling model of each token is optimized by minimizing a Kullback-Leibler divergence (KLD) objective. We demonstrate that the CTC-weighted language model log likelihood ratio can be interpreted as the KLD between the unknown correct ASR and the unsupervised learned ASR, and that, using a Pinsker bound, the square root of KLD can be interpreted as an upper bound on the total variation distance between the true and estimated spelling of the unknown word. Experiments show relative OOV character-error-rate reductions of up to 14.97$\%$ on LibriSpeech and 6.67$\%$ on dysarthric Speech Accessibility Project data for recurring OOV words, relative to the corresponding rescoring system.

\end{abstract}

\section{Introduction}

It is possible for modern automatic speech recognition (ASR) to
correctly recognize a word it has never seen before if it can
correctly recognize the sequence of sub-word units, e.g., characters
or byte-pair encoded (BPE) wordparts~\cite{sennrich2016neural}.
Unknown words tend to be misunderstood by ASR, however, in part
because the ASR lacks an adequate model of the word's usage contexts,
and
therefore cannot justify its transcription using a language model
(LM)~\cite{pundak2018deep}.  
First-language~\cite{christ2018hearing} and second-language
acquisition researchers~\cite{webb2019incidental} distinguish between intentional
vocabulary (learned in formal settings) and incidental vocabulary
(learned from informal conversation).  It is desirable that a
human-like agent should be capable, as humans are, of learning new
words incidentally during informal conversation, then adding the new
words to its LM for use in future conversations.
This task is markedly different from most studies of OOV recognition in the literature,
in which the spelling of the OOV word is often provided, but its pronunciation is unknown.
When the spelling is available, it can be provided to the ASR in a context biasing
list~\cite{sathyendra2022contextual,bruguier2019phoebe,dingliwal2023personalization},
or as prompt or RAG information to a speech LLM~\cite{abouelenin2025phi,zhang2024internlm,zeng2024glm,xie2024mini,chu2024qwen2} or deep fusion system~\cite{le2021deep,pundak2018deep}.  When the spelling of all OOV words
is as unknown as their pronunciation, context biasing methods are less effective. Prior works recover OOV words from hypothesized phoneme sequences by phoneme-to-grapheme (P2G) conversion, followed by lexicon integration and n-gram LM score estimation \cite{qin2014building}, or by RNNLM vocabulary expansion with representations from learned lexical features \cite{zhang2020oov}. In contrast, this paper proposes to correctly transcribe words whose spelling and pronunciation have never been seen before by test-time lexical adaptation, which detects candidate lexical mismatches, initializes new neural-LM tokens with lexical identities generated by Connectionist Temporal Classification (CTC,~\cite{graves2006connectionist}), and updates only the new token representations and spelling distributions from unlabeled speech data.

ASR maps from a sequence of speech,
$\mathbf{x}_{0:T}=[x_0,\ldots,x_{T-1}]$, $x_t\in\mathcal{X}$, to a sequence of
LM tokens, $\mathbf{y}_{0:L}=[y_0,\ldots,y_{L-1}]$, $y_l\in\mathcal{Y}$.
Training efficiency is improved by imposing monotonicity and
compressivity as modeling constraints, e.g., using the translation and compression stages of CTC.  
The translation stage generates a character
sequence $\mathbf{c}_{0:T}=[c_0,\ldots,c_{T-1}]$, where each character is an LM
token, part of an LM token, or a null character.
The compression stage generates $\mathbf{y}_{0:L}$ using a deterministic
algorithm, $\mathbf{y}=\mathcal{B}(\mathbf{c})$,
and $p(c|\mathbf{x})$ is trained to maximize
the log probability of the training dataset:
\begin{align}
  \log p(\mathbf{y}|\mathbf{x}) &= \log\sum_{\mathbf{c}\in\mathcal{B}^{-1}(\mathbf{y})} 
  \prod_{t=0}^{T-1} p(c_t|\mathbf{x}_{0:T})
\end{align}



Text corpora tend to have several orders of magnitude more data
(measured in LM token count) than speech corpora, and perhaps for this
reason, the error rate of an ASR can be reduced using shallow fusion
with an LM~\cite{kannan2018analysis}.  Shallow fusion chooses the
transcription with maximum fusion score, $f(\mathbf{x},\mathbf{y})$, defined as
\begin{equation}
  f(\mathbf{x},\mathbf{y}) = \log p(\mathbf{y})+\log\sum_{\mathbf{c}\in\mathcal{B}^{-1}(\mathbf{y})} p(\mathbf{c}|\mathbf{x})+\gamma L,
  \label{eq:asr_score}
\end{equation}
where $\gamma L$ is a length penalty proportional to the transcript
length $L$, and $p(\mathbf{y})$ is an LM.
The effectiveness of shallow fusion can be theoretically analyzed by
introducing an explicit spelling model, $p(\mathbf{c}|\mathbf{y})$.  
Using Eq.~(10) from Morgan and
Bourlard~\cite{morgan1995continuous}, it is possible to show that
\begin{equation}
  \log p(\mathbf{y}|\mathbf{x}) = \log \left[
  p(\mathbf{y})\sum_{\mathbf{c}\in\mathcal{B}^{-1}(\mathbf{y})} p(\mathbf{c}|\mathbf{y})\frac{p(\mathbf{c}|\mathbf{x})}{p(\mathbf{c})}
  \right].
  \label{eq:morgan}
\end{equation}
Eq.~(\ref{eq:morgan}) is equal to Eq.~(\ref{eq:asr_score}) if
$p(\mathbf{c}|\mathbf{y})=p(\mathbf{c})e^{\gamma L}$ for all $\mathbf{c}\in\mathcal{B}^{-1}(\mathbf{y})$, i.e., if
all permissible spellings of a word have conditional probabilities proportional to
their marginals.
Instead, this paper proposes to model
$p(\mathbf{c}|\mathbf{y})$ and $p(\mathbf{c})$ using explicit spelling models.

\section{Unsupervised Learning of New Words}

The LM token vocabulary is made up of sub-word
units, e.g., units computed by byte-pair encoding (BPE) a large
training corpus~\cite{gage1994new}.  Though BPE units are designed to
represent sub-words, a sufficiently large BPE
vocabulary represents each of the most common words in the training
corpus with its own language model token~\cite{chung2026exploiting}.  
Experimental results with five different large language models~\cite{tao2024scaling}
showed that 
optimal language modeling performance is achieved if vocabulary size scales as the $0.84^{\text{th}}$ power of the number of
non-vocabulary parameters~\cite{tao2024scaling}.
Similar experimental results with a different set of language models~\cite{kaplan2020scaling}
showed that non-vocabulary parameter count 
should scale in proportion to the $1.35^{\text{th}}$ power of the
amount of training data, therefore representing a new
word by simply increasing the size of the BPE token vocabulary may be
a reasonable approach.

Let $\mathbf{x}_{s:t}=[x_s,\ldots,x_{t-1}]$ be an acoustic segment for which
the ASR has low confidence.
More precisely, suppose
there is no known LM token $y_{l}$ with spelling $\mathbf{c}_{s:t}$
that has both a high language-model probability 
$p(\mathbf{y}_{0:l},y_l,\mathbf{y}_{l+1:L})$
and a high acoustic
model probability
$\sum_{\mathbf{c}\in c(y_l)}p(\mathbf{c}|y_l)p(\mathbf{c}|\mathbf{x}_{s:t})/p(\mathbf{c})$.
Incidental word learning is achieved by expanding the
vocabulary of the language model to
$\mathcal{Y}\leftarrow\mathcal{Y}\cup\{\hat{y}\}$, where the meaning
of the new word is chosen to maximize $p(\mathbf{y})$.
A new vocabulary item $\hat{y}$ can be added to a Transformer LLM\cite{vaswani2017attention}, minimally, by initializing a new output weight vector
$v_{\hat{y}}$ and input embedding vector
$e_{\hat{y}}$, then adjusting them, together with
spelling model parameters, to optimize some unsupervised training criterion.


Suppose that the spelling of the new word is unknown 
and must be estimated.
Suppose that the resources available for doing so include a set of unlabeled 
speech samples, $\mathcal{D}=\{\mathbf{x}^{(0)},\ldots,\mathbf{x}^{(D-1)}\}$, 
at least one of which is believed to contain $\hat{y}$, and 
a well-trained acoustic model $p(\mathbf{c}|\mathbf{x})$.
With these resources, we can estimate $p(\mathbf{c})$, the marginal probability of any particular
sequence of characters, as
\begin{align}
  p(\mathbf{c})=\sum_{\mathbf{x}} p(\mathbf{x})p(\mathbf{c}|\mathbf{x})&\underset{|\mathcal{D}|\rightarrow\infty}{\longleftarrow}
  \frac{1}{|\mathcal{D}|}\sum_{\mathbf{x}\in\mathcal{D}} p(\mathbf{c}|\mathbf{x}).
\end{align}
The spelling model $q(\mathbf{c}|\hat{y})$ can be estimated by
minimizing the Kullback-Leibler divergence (KLD) between its marginal, $q(\mathbf{c})=\sum_y q(\mathbf{y})q(\mathbf{c}|\mathbf{y})$, and
the known $p(\mathbf{c})$:
\begin{align}
  &D(p(\mathbf{c})\Vert q(\mathbf{c})) = -\sum_{\mathbf{c}} p(\mathbf{c})\log\left(\frac{q(\mathbf{c})}{p(\mathbf{c})}\right)
  \label{eq:kld}\\
  &  \underset{|\mathcal{D}|\rightarrow\infty}{\longleftarrow}
  -\frac{1}{|\mathcal{D}|}\sum_{\mathbf{x}\in\mathcal{D}}\sum_{\mathbf{c}} p(\mathbf{c}|\mathbf{x})
  \log\left(\frac{\sum_{\mathbf{y}} q(\mathbf{c}|\mathbf{y})q(\mathbf{y})}{p(\mathbf{c})}\right).\label{eq:lnw}
\end{align}

If there is no way to determine that the same new word has occurred in
multiple test utterances, then $D(p\Vert q)$ can be estimated using
only one example, at the cost of increased uncertainty in the
resulting estimated spelling model $q(\mathbf{c}|\mathbf{y})$.  On the other hand, given a
sufficiently large number of unlabeled examples,
if the unknown word are available, 
Pinsker's inequality~\cite{pinsker1960information}
guarantees that the total variation distance (TVD) between $p(\mathbf{c})$ and
$q(\mathbf{c})$ is bounded by
\begin{align}
 \sqrt{2 D(p\Vert q)} \ge \sum_{\mathbf{c}} \vert p(\mathbf{c})-q(\mathbf{c})\vert.
 \label{eq:pinsker4}
\end{align}
Both $D(p\Vert q)$ and $\text{TVD}(p-q)=\sum\vert{p(\mathbf{c})-q(\mathbf{c})}\vert$ are convex functions
of $q(\mathbf{c})$, and both have a single global minimum at the value $q(\mathbf{c})=p(\mathbf{c})$.
Convergence to zero of $D(p\Vert q)$ therefore guarantees pointwise
convergence of $q(\mathbf{c})$ to $p(\mathbf{c})$ for every character sequence $\mathbf{c}$.

\begin{figure*}[t]
    \centering
    \vspace{-1.5em}
    \resizebox{0.83\textwidth}{!}{
    \includegraphics[width=\textwidth]{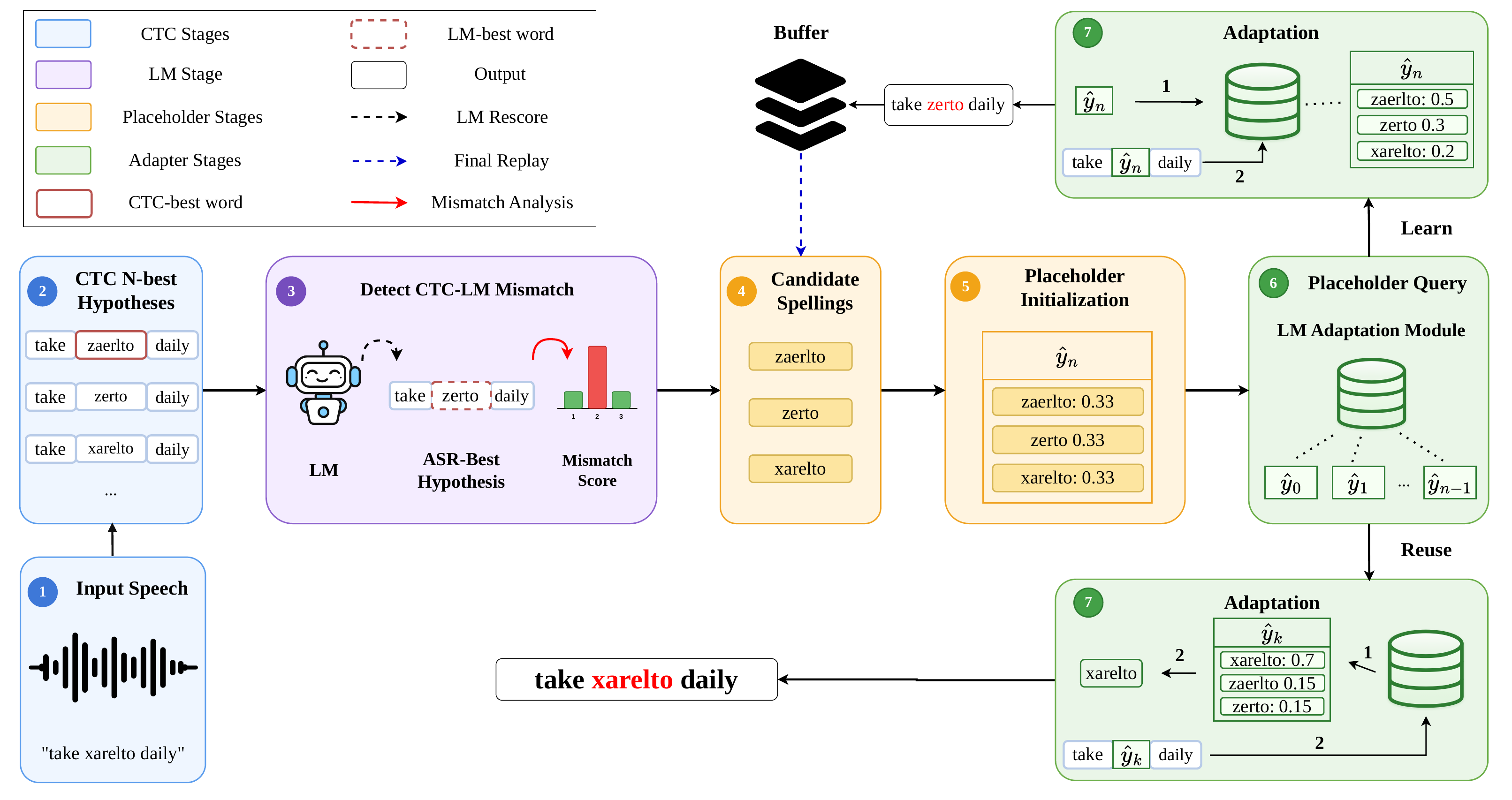}
    }
    \vspace{-1.0em}
    \caption{Overview of proposed test-time adaptation framework. The CTC model generates N-best hypotheses for an input utterance. Word spans with strong CTC-LM preference disagreement are selected as potential OOV regions (one span here). Top-K CTC spellings are used to initialize a placeholder token with uniform spelling prior. For the first occurrence, the placeholder is added into the vocabulary and fine-tuned in context, while the utterance is buffered for potential final refinement. Otherwise, we retrieve the existing placeholder and re-transcribe the test utterance, possibly reducing errors. 
    The retrieved placeholder is further updated using unsupervised model adaptation. 
    }
    \label{fig:workflow}
    \vspace{-1.5em}
\end{figure*}

\section{Overall Architecture}
Our system is demonstrated in Fig.~\ref{fig:workflow}.
The HuBERT-XLarge CTC model maps the input speech to frame-level acoustic logits, from which a CTC beam search decoder generates N-best hypotheses with corresponding acoustic scores. To locate the acoustic segment $\mathbf{x}_{s:t}$, we utilize the 1-best hypothesis as the anchor and align the remaining hypotheses to it. For each word span in the anchor hypothesis, we collect the corresponding alternatives proposed by CTC and score them with the LM under the current context, which allows us to estimate the language model preference among acoustically plausible candidates and identify those spans where the acoustic evidence and language model preference disagree. To quantify this mismatch, we define a CTC-LM mismatch score as follows.

For the selected span $\mathbf{x}_{s:t}$, we assume $\mathcal{Y}_{s:t}$ denotes the set of proposed CTC alternatives for it. We define
\begin{align}
&
    y_l^{\mathrm{ctc}} = \argmax_{y \in \mathcal{Y}_{s:t}} \log p_{ctc}(y | \mathbf{x}_{s:t})~\text{and}~ \\
&
    y_l^{\mathrm{lm}} = \argmax_{y \in \mathcal{Y}_{s:t}} \log p_{lm}( \mathbf{y}_{0:l}, y, \mathbf{y}_{(l+1):L})
\end{align}
as preferred candidates for CTC model and language model, respectively. Then we define the mismatch score $M(\mathbf{x}_{s:t})$ as
    $M(\mathbf{x}_{s:t}) = \Delta_{ctc} + \lambda_{lm} \Delta_{lm}$,
where the coefficient $\lambda_{lm}$ is a tunable weight that controls the contribution of the language model gap, and 
\begin{align*}
&
    \Delta_{ctc} = \log p_{ctc}(y_l^{\mathrm{ctc}} | \mathbf{x}_{s:t}) - \log p_{ctc}(y_l^\mathrm{lm} | \mathbf{x}_{s:t}), \\
&
    \Delta_{lm} = \log p_{lm}(  \mathbf{y}_{0:l},y_l^\mathrm{lm},\mathbf{y}_{(l+1):L}) - \log p_{lm}(\mathbf{y}_{0:l},y_l^\mathrm{ctc},\mathbf{y}_{(l+1):L}).
\end{align*}

The LM token vocabulary is expanded iteratively. In each iteration, the test span with the highest mismatch score is selected to be evaluated as the possible locus of a new LM token, $\hat{y}_n$, where $n$ is the number of OOV tokens that have already been created from previous unlabeled test utterances.  All compressed character sequences in the CTC beam are re-aligned to the CTC 1-best time alignment, and to the LM token sequence $[\mathbf{y}_{0:l},\hat{y}_n,\mathbf{y}_{l+1:L}]$, and from these alignments, the top-K candidate spellings are extracted.  The new word's spelling model is initialized uniformly, $q(\mathbf{c}|\hat{y}_n)=\frac{1}{K}$ for each of the top-K spellings.  The new LM token is then retained for future use only if its spelling model is different from the spelling models of any previously discovered OOV token $\hat{y}_k$ for $k<n$, specifically, only if:
\begin{align*}
    \min_{\mathbf{c}_n \in \mathcal{B}^{-1}(\hat{y}_{n}), \mathbf{c}_k \in \mathcal{B}^{-1}(\hat{y}_{k})}
    \Big(
    \frac{D_{\text{edit}}(\mathbf{c}_n,\mathbf{c}_k)}{\max(|\mathbf{c}_n|, |\mathbf{c}_k|)} 
    \Big) \geq \delta, \indent \forall k < n,
\end{align*}
where $D_{\text{edit}}$ is the Levenshtein edit distance, and $\delta$ is the reuse threshold.
If no placeholder can be retrieved,  $\hat{y}_{n}$ is added into the vocabulary of the adaptation module. We exclusively fine-tune the input embedding and output softmax parameters associated with $\hat{y}_{n}$ in the current context to minimize Eq. (\ref{eq:lnw}), while keeping all other parameters frozen. Since $\hat{y}_{n}$ is optimized using the same utterance where it is initially detected, it will be stored in the registry and the utterance will be buffered and rescored in the end if the placeholder has accumulated more evidence from subsequent occurrences, thereby reducing self-adaptation bias.


 If an existing placeholder $\hat{y}_{k}$ can be retrieved from the placeholder registry, we first determine whether any of its candidate spellings, $\mathbf{c}_{s:t}\in \mathcal{B}^{-1}(\hat{y}_k)$, explains the current context better than the known spelling $\mathbf{c}(y_l)$ of the original span $y_l$. The decision is made using a local score: 
\begin{align*}
\resizebox{0.99\columnwidth}{!}{$
\begin{aligned}
S(\hat{y}_{k}, y_l)
&= \alpha_{\mathrm{lm}}
\big[
\log q(\mathbf{y}_{0:l}, \hat{y}_{k}, \mathbf{y}_{(l+1):L})
-
\log q(\mathbf{y}_{0:l}, y_l, \mathbf{y}_{(l+1):L})
\big] \\
&\quad+
\big[
\max_{\mathbf{c}_{s:t} \in \mathcal{B}^{-1}(\hat{y}_{k})}
\log p(\mathbf{c}_{s:t} \mid \mathbf{x}_{s:t})
-
\log p(\mathbf{c}(y_l) \mid \mathbf{x}_{s:t})
\big],
\end{aligned}
$}
\end{align*}
 where $\alpha_{lm}$ is a tunable weight and we only accept replacement when $S(\hat{y}_{k}, y_l) > 0$ and the substituted placeholder yields improvements in both the contextual LM likelihood and the local acoustic evidence over the original span.
 If the replacement is accepted, then both the recognized transcription of the test utterance and the parameters of $\hat{y}_k$ are updated.  The recognized transcription, $\mathbf{c}^{\star}$, is
\begin{align*}
\mathbf{c}^\star 
&= \argmax_{\mathbf{c}_{s:t} \in \mathcal{B}^{-1}(\hat{y}_{k})}
\Big[
\log p(\mathbf{c}_{s:t} \mid \mathbf{x}_{s:t})
+ \lambda_p \log q(\mathbf{c}_{s:t} \mid \hat{y}_{k})
\Big],
\end{align*}
where $q(\mathbf{c}_{s:t}|\hat{y}_{k})$ is the spelling model learned from previous test utterances, and $\lambda_{p}$ is a tunable hyperparameter.  In parallel, the spelling model $q(\mathbf{c}_{s:t}|\hat{y}_{k})$, embedding $e_{\hat{y}_k}$, and softmax weights $v_{\hat{y}_k}$ are adapted using 100 steps of gradient descent on Eq.~(\ref{eq:lnw}), allowing the model to incorporate new acoustic and contextual evidence across multiple occurrences.

\begin{table*}[t]
\centering
\vspace{-0.5em}
\caption{Results on LibriSpeech (\%). ``OOV subset'' denotes the set of test utterances that contain frequent OOV words. Parentheses indicate the relative error reduction (\%) over the corresponding shallow fusion system. Asterisks denote statistically significant differences from the corresponding plain rescoring baselines: * denotes significance at $p<0.05$ and ** at $p\leq0.001$, measured with the NIST \texttt{sc\_stats} tool.}
\label{tab:wer/cer_ls}

\setlength{\tabcolsep}{1.5pt}
\renewcommand{\arraystretch}{0.7}
\vspace{0.3em}

\begin{tabular}{c|c|ccccccc}
\toprule

\multicolumn{2}{c|}{\textbf{Datasets}}
&
\makecell{\textbf{HuBERT-XLarge}\\\textbf{(CTC)}}
&
\makecell{\textbf{GPT-2}\\\textbf{Rescoring}}

&
\makecell{\textbf{GPT-2}\\\textbf{+ Adapter}}
&
\makecell{\textbf{Qwen-3}\\\textbf{Rescoring}}
&
\makecell{\textbf{Qwen-3}\\\textbf{+ Adapter}}
&
\makecell{\textbf{Gemma-3}\\\textbf{Rescoring}}
&
\makecell{\textbf{Gemma-3}\\\textbf{+ Adapter}}
\\

\midrule

\renewcommand{\arraystretch}{0.7}

\multirow[c]{2}{*}[0.5ex]{%
  \makecell[c]{\textbf{test-clean}\\\textbf{overall}}
}
& \textbf{WER}
& 1.83 & 1.91 & 1.89(1.05) & 1.77 & 1.77 (0.00) & 1.80 & 1.79 (0.56) \\[1pt]

& \textbf{CER}
& 0.52 & 0.55 & 0.54 (1.82) & 0.51 & 0.51 (0.00) & 0.51 & 0.51 (0.00) \\

\midrule

\multirow[c]{2}{*}[0.5ex]{%
  \makecell[c]{\textbf{test-other}\\\textbf{overall}}
}
& \textbf{WER}
& 3.39 & 3.49 & 3.41 (2.29**) & 3.24 & 3.21 (0.93) & 3.25 & 3.23 (0.62) \\[1pt]

& \textbf{CER}
& 1.05 & 1.10 & 1.07 (2.73**) & 1.03 & 1.02 (0.97) & 1.04 & 1.03 (0.96)\\

\midrule

\multirow[c]{3}{*}[-0.3ex]{\makecell{\textbf{test-clean}\\\textbf{OOV subset}}}
& \textbf{WER}
& 5.23 & 5.09 & 4.92 (3.34) & 4.95 & 4.81 (2.83) & 4.88 & 4.63 (5.12*) \\[1pt]

& \textbf{CER}
& 1.66 & 1.69 & 1.55 (8.28**) & 1.67 & 1.57 (5.99*) & 1.60 & 1.51 (5.63**) \\[1pt]

& \textbf{OOV-CER}
& 18.05 & 18.54 & 16.47 (11.17**) & 18.05 & 16.27 (9.86*) 
& 17.85 & 16.07 (9.97*) \\
\midrule

\multirow[c]{3}{*}[-0.3ex]{\makecell{\textbf{test-other}\\\textbf{OOV subset}}}
& \textbf{WER}
& 7.87 & 7.97 & 7.50 (5.90*) & 7.67 & 7.34 (4.30*) & 8.00 & 7.60 (5.00*) \\ [1pt]

& \textbf{CER}
& 2.30 & 2.45 & 2.25 (8.16**) & 2.32 & 2.19 (5.60**) & 2.34 & 2.24 (4.27*) \\ [1pt]

& \textbf{OOV-CER}
& 17.77 & 20.24 & 17.21 (\textbf{14.97**}) & 20.16 & 17.53 (\textbf{13.05*}) & 21.04 & 18.25 (\textbf{13.26*}) \\

\bottomrule
\end{tabular}
\vspace{-2.1em}
\end{table*}

\section{Experimental Settings}
Simulations with speech data were conducted using the HuBERT-XLarge-LS960-FT CTC
model\cite{hsu2021hubert} as a pre-trained acoustic model ($p(\mathbf{c}|\mathbf{x})$), 
the GPT-2 \cite{radford2019language}, Qwen-3-0.6B-Base \cite{yang2025qwen3}, and Gemma-3-270M \cite{team2025gemma} models of all known words ($p(\mathbf{y})$), which remain frozen throughout inference, and by modifying them using a dynamic OOV adaptation module to
create the models of new words ($q(\mathbf{y})$). HuBERT-XLarge CTC provides strong frame-level speech representations and a simple CTC decoding interface.
We choose GPT-2, Qwen-3-0.6B-Base, and Gemma-3-270M as
the external language models because they employ competitive language
modeling capability while preserving moderate computational overhead,
making them preferable for test-time adaptation. More importantly, all architectures support direct vocabulary expansion and modification on the embedding layers and output softmax layers, allowing new lexical items to be added and fine-tuned exclusively.
For each language model, the corresponding adaptation module is instantiated from the same model, which is responsible for dynamic test-time OOV adaptation. The placeholder is initialized with top-K CTC spellings, and we report the results using K=5.


We evaluate our system with two types of speech data. First, we use the LibriSpeech dataset\cite{panayotov2015librispeech} which provides a standard speech benchmark for evaluating general ASR performance. Since the HuBERT-XLarge model has already been fine-tuned on LibriSpeech, we only utilize test-clean and test-other sets for evaluation. Second, we employ a more challenging  dysarthric speech dataset from the 2026-04-30 release of Speech Accessibility Project (SAP\cite{hasegawa2024community}), which contains recordings from speakers with speech disabilities. For this setting, we fine-tune HuBERT-XLarge on the SAP training set, which contains approximately 740 hours of speech, and test our system on the SAP development set.

In our experiments, a word is considered OOV if it is in the evaluation set but absent from all supervised ASR training sets for the CTC model. Importantly, no target OOV identity, spelling, or candidate list is provided at test time. Results are reported only for OOV words that occur more than once, and the test utterances that contain them, so that a placeholder adapted from an initial occurrence can be applied to subsequent instances. Hyperparameters are tuned on the frequent-OOV subsets derived from the development sets and kept fixed for all experiments on the corresponding test sets with the same language model. In LibriSpeech, there are 123 utterances in the test-clean and 135 utterances in the test-other satisfying the requirement. In the SAP development set, we identify 921 utterances with frequent OOV words that do not appear in the SAP training set or the LibriSpeech training sets. We reserve 200 of them for hyperparameter tuning and evaluate on the remaining 721 utterances.


\begin{table}[t]
\centering
\vspace{-0.5em}
\caption{Results on SAP-dev OOV subset (\%). Parentheses show relative error reduction (\%) over the corresponding shallow fusion system. Significant reductions: * for $p<0.05$ and ** for $p\leq0.001$.}
\label{tab:wer/cer_sap}

\normalsize
\setlength{\tabcolsep}{1.2pt}
\renewcommand{\arraystretch}{0.85}
\vspace{0.1em}

\begin{tabular}{@{}l@{\hspace{2.5pt}}|ccc@{}}
\toprule
\multirow{2}{*}{\textbf{System}}

& \multicolumn{3}{c}{\textbf{SAP-dev OOV subset}} \\
\cmidrule(lr){2-4}
& \textbf{WER} & \textbf{CER} & \textbf{OOV-CER} \\

\midrule

HuBERT-XLarge
& 28.96 & 15.94 & 40.64 \\

\midrule

\quad + GPT-2 
& 26.41 & 14.35 & 38.97 \\

\qquad + \textit{Adapter} 
& 25.70 (2.69**) & 13.85 (3.48**) & 36.37 (\textbf{6.67**}) \\[3pt]



\quad + Qwen-3
& 25.53 & 14.03 & 38.21 \\

\qquad + \textit{Adapter} 
& 25.23 (1.18*) & 13.75 (2.00**) & 36.58 (\textbf{4.27**}) \\[3pt]

\quad + Gemma-3
& 25.97 & 13.96 & 38.21 \\

\qquad + \textit{Adapter}
& 25.49 (1.85**) & 13.63 (2.36**) & 36.74 (\textbf{3.85**}) \\

\bottomrule
\end{tabular}
\vspace{-2.5em}
\end{table}

\section{Results}

Table~\ref{tab:wer/cer_ls} summarizes the results on the LibriSpeech benchmark. For each test set, we evaluate the performance in two settings: the complete test set and the subset of utterances containing frequent OOV words. In the full test sets, the shallow fusion of HuBERT-XLarge+GPT-2 slightly increases WER/CER relative to the standalone CTC system, while the stronger Qwen-3 and Gemma-3 yield modest improvements. The effect of test-time adaptation is small: it produces marginal gains over plain rescoring for GPT-2 and Gemma-3 systems in both test sets, while no improvement is observed for the Qwen-3 system in the test-clean set. The limited overall improvement is expected since OOV words constitute only a small portion of the complete test sets, resulting in a diluted effect in the overall evaluation.

In the OOV subsets, we additionally examine the CER of OOV words separately.
The shallow fusion with GPT-2 and Gemma-3 slightly degrades WER/CER relative to the HuBERT-XLarge alone on test-other, and the rescoring systems generally increase the OOV-CER. In contrast, the proposed adaptation systems consistently reduce the WER/CER/OOV-CER compared with the corresponding base systems across all language models. Paired significance tests with the NIST \texttt{sc\_stats} tool \cite{pallet1990tools} show that most improvements are either statistically significant ($p < 0.05$) or highly statistically significant ($p \leq 0.001$). Specifically, we observe a relative OOV-CER reduction of up to 14.97$\%$, with the largest reduction being highly statistically significant. These results demonstrate that the proposed adaptation generalizes across different language model backbones and improves the recognition of unseen words.

In Table~\ref{tab:wer/cer_sap}, we present the WER, CER, and OOV-CER results on 
the OOV subset of the SAP development set. In this challenging dysarthric speech dataset, the shallow fusion with all three language models consistently outperforms the HuBERT-XLarge baseline, perhaps because the intelligibility degradations caused by dysarthria are partially mitigated by the use of LM. With the adaptation modules, we observe small but significant error reductions relative to the corresponding shallow fusion systems. The largest gains are observed in OOV-CER, with relative reductions of 6.67$\%$, 4.27$\%$ and 3.85$\%$ for GPT-2, Qwen-3, and Gemma-3, respectively.

\section{Discussion}
Experimental results demonstrate the feasibility of learning new LM tokens from unlabeled speech and applying the learned information to subsequent encounters. The proposed adaptation system can significantly improve the transcription of OOV words in both clean and degraded speech conditions. We discover that adaptation performance is strongly affected by two factors: (1) the accurate detection of OOV spans and (2) the reliable generation of acoustically plausible CTC spellings. In some failure cases, the OOV detection is confounded by acoustically ambiguous segments or rare in-vocabulary words, which may prevent the system from targeting true OOV spans and result in erroneous adaptation. Moreover, if CTC confidently misrecognizes the OOV spans, the adaptation process will not be triggered due to the lack of CTC-LM disagreement. Even when the OOV spans are correctly localized, the quality of CTC-proposed spellings remains a bottleneck, since the adaptation module can only propose new-word spellings based on the set of options provided by CTC from the first occurrence. If the correct spelling is absent from the top-K spellings, or the dominant spelling in the placeholder is unreliable, subsequent placeholder reuse might introduce more errors into the hypotheses. Therefore, future work should investigate more robust OOV detection mechanisms and test-time strategies to dynamically optimize the spellings in the placeholder.

The adaptation strategy in our setting is conservative: when an OOV word is encountered for the first time, the system learns the spellings and buffers the corresponding utterance for potential revision if the subsequent occurrences update the placeholder under different contexts; otherwise, we preserve the original hypothesis. This design reduces the risk of self-adaptation bias, but limits the benefit on the OOV words that occur only once. Future research is necessary to explore more advanced adaptation strategies that expand the revision coverage while preserving the adaptation reliability.

\section{Conclusion}

Intelligent behavior includes the ability to learn new words
incidentally, from their use in context.  An ASR can learn incidental words by adding a new token to the BPE
vocabulary of the language model, estimating its contextual representation from the
language model context, and estimating its spelling using the top-K
character sequences produced by a CTC acoustic model.  
Unsupervised estimation
of new word models permits the ASR to recognize future instances of
the same word with reduced error rate.

\begingroup
\bibliographystyle{IEEEbib}
\bibliography{strings,refs}

@string{acl = {Proceedings of the 61st Annual Meeting of the Association for Computational Linguistics (Volume 1: Long Papers)}}

@string{icassp = {IEEE International Conference on Acoustics, Speech and Signal Processing (ICASSP)}}

@string{icml = "Proc. International Conference on Machine Learning (ICML)"}

@string{spm = "{IEEE} Signal Processing Magazine"}

@article{abouelenin2025phi,
  title={Phi-4-mini technical report: Compact yet powerful multimodal language models via mixture-of-loras},
  author={A. Abouelenin and A. Ashfaq and A. Atkinson and H. Awadalla and N. Bach and J. Bao and A. Benhaim and M. Cai and V. Chaudhary and C. Chen and others},
  journal={arXiv:2503.01743},
  year={2025}
}

@inproceedings{bruguier2019phoebe,
  title={Phoebe: Pronunciation-aware contextualization for end-to-end speech recognition},
  author={A. Bruguier and R. Prabhavalkar and G. Pundak and T. N. Sainath},
  booktitle={Proc. ICASSP},
  address={Brighton},
  pages={6171--6175},
  year={2019},
}

@article{christ2018hearing,
  title={Hearing words, learning words: How different presentations of novel vocabulary words affect children’s incidental learning},
  author={T. Christ and M. M. Chiu},
  journal={Early Education and Development},
  volume={29},
  number={6},
  pages={831--851},
  year={2018},
  publisher={Taylor \& Francis}
}

@article{chu2024qwen2,
  title={{Qwen2-Audio} technical report},
  author={Y. Chu and J. Xu and Q. Yang and H. Wei and X. Wei and Z. Guo and Y. Leng and Y. Lv and J. He and J. Lin and others},
  journal={arXiv:2407.10759},
  year={2024}
}

@article{chung2026exploiting,
  title={Exploiting vocabulary frequency imbalance in language model pre-training},
  author={W. Chung and J. Kim},
  journal={Advances in Neural Information Processing Systems},
  volume={38},
  pages={137676--137706},
  year={2026}
}

@inproceedings{dingliwal2023personalization,
  title={Personalization of {CTC} speech recognition models},
  author={S. Dingliwal and M. Sunkara and S. Ronanki and J. Farris and K. Kirchhoff and S. Bodapati},
  booktitle={Proc. SLT},
  address={Doha},
  year={2023},
  pages={302--309}
}

@article{gage1994new,
  title={A new algorithm for data compression},
  author={P. Gage},
  journal={C Users Journal},
  volume={12},
  number={2},
  pages={23--38},
  year={1994},
  publisher={McPherson, KS: R \& D Publications, c1987-1994.},
  }

@inproceedings{graves2006connectionist,
 title={Connectionist Temporal Classification: Labelling Unsegmented Sequence Data with Recurrent Neural Networks},
 author={A. Graves and S. Fern{\'{a}}ndez and F. Gomez and J. Schmidhuber},
 booktitle={Proc. ICML},
 address={Pittsburgh},
 year={2006},
 pages={369--376}
}

@article{hasegawa2024community,
  title={Community-supported shared infrastructure in support of speech accessibility},
  author={M. Hasegawa-Johnson and X. Zheng and H. Kim and C. Mendes and M. Dickinson and E. Hege and C. Zwilling and M. M. Channell and L. Mattie and H. Hodges and others},
  journal={Journal of Speech, Language, and Hearing Research},
  volume={67},
  number={11},
  pages={4162--4175},
  year={2024},
  publisher={American Speech-Language-Hearing Association}
}

@article{hsu2021hubert,
  title={{HuBERT}: Self-supervised speech representation learning by masked prediction of hidden units},
  author={W.-N. Hsu and B. Bolte and Y.-H. H. Tsai and K. Lakhotia and R. Salakhutdinov and A. Mohamed},
  journal={IEEE/ACM transactions on audio, speech, and language processing},
  volume={29},
  pages={3451--3460},
  year={2021},
  publisher={IEEE}
}

@inproceedings{kannan2018analysis,
  title={An analysis of incorporating an external language model into a sequence-to-sequence model},
  author={A. Kannan and Y. Wu and P. Nguyen and T. N. Sainath and Z. Chen and R. Prabhavalkar},
  booktitle={Proc. ICASSP},
  address={Calgary},
  year={2018},
  pages={1-5828},
}

@article{kaplan2020scaling,
  title={Scaling laws for neural language models},
  author={J. Kaplan and S. McCandlish and T. Henighan and T. B. Brown and B. Chess and R. Child and S. Gray and A. Radford and J. Wu and D. Amodei},
  journal={arXiv:2001.08361},
  year={2020}
}

@inproceedings{le2021deep,
  title={Deep shallow fusion for {RNN-T} personalization},
  author={D. Le and G. Keren and J. Chan and J. Mahadeokar and C. Fuegen and M. L. Seltzer},
  booktitle={Proc. SLT},
  address={Shenzhen},
  year={2021},
  pages={251--257},
}

@article{morgan1995continuous,
 author="N. Morgan and H. Bourlard",
 journal=spm,
 number="3",
 pages="24-42",
 title="Continuous Speech Recognition",
 volume="12",
 year="1995"
}

@inproceedings{pallet1990tools,
 title={Tools for the analysis of benchmark speech recognition tests},
 author={D. S. Pallet and W. M. Fisher and J. G. Fiscus},
 booktitle={Proc. ICASSP},
 address={Albuquerque},
 year={1990},
 pages={97--100},
}

@inproceedings{panayotov2015librispeech,
  title={{LibriSpeech}: an {ASR} corpus based on public domain audio books},
  author={V. Panayotov and G. Chen and D. Povey and S. Khudanpur},
  booktitle={Proc. ICASSP},
  address={Brisbane},
  year={2015},
  pages={5206--5210}
}

@book{pinsker1960information,
 author={M. S. Pinsker},
 title={Information and Information Stability of Random Variables and Processes (in Russian)},
 address={Moscow},
 publisher={Izv. Akad. Nauk},
 year={1960}
}

@inproceedings{pundak2018deep,
  title={Deep context: end-to-end contextual speech recognition},
  author={G. Pundak and T. N. Sainath and R. Prabhavalkar and A. Kannan and D. Zhao},
  booktitle={Proc. SLT},
  address={Athens},
  year={2018},
  pages={418--425}
}

@article{radford2019language,
  title={Language models are unsupervised multitask learners},
  author={A. Radford and J. Wu and R. Child and D. Luan and D. Amodei and I. Sutskever and others}, 
  journal={OpenAI blog},
  volume={1},
  number={8},
  pages={9},
  year={2019}
}

@article{vaswani2017attention,
  title={Attention is all you need},
  author={A. Vaswani and N. Shazeer and N. Parmar and J. Uszkoreit and L. Jones and A. N. Gomez and {\L}. Kaiser and I. Polosukhin},
  journal={Advances in neural information processing systems},
  volume={30},
  year={2017}
}

@inproceedings{sathyendra2022contextual,
  title={Contextual adapters for personalized speech recognition in neural transducers},
  author={K. M. Sathyendra and T. Muniyappa and F.-J. Chang and J. Liu and J. Su and G. P. Strimel and A. Mouchtaris and S. Kunzmann},
  booktitle={Proc. ICASSP},
  address={Singapore},
  year={2022},
  pages={8537--8541}
}

@inproceedings{sennrich2016neural,
  title={Neural machine translation of rare words with subword units},
  author={R. Sennrich and B. Haddow and A. Birch},
  booktitle={Proc. ACL},
  address={Berlin},
  year={2016},
  pages={1715--1725}
}

@article{tao2024scaling,
  title={Scaling laws with vocabulary: Larger models deserve larger vocabularies},
  author={C. Tao and Q. Liu and L. Dou and N. Muennighoff and Z. Wan and P. Luo and M. Lin and N. Wong},
  journal={Advances in Neural Information Processing Systems},
  volume={37},
  pages={114147--114179},
  year={2024}
}

@incollection{webb2019incidental,
  title={Incidental vocabulary learning},
  author={S. Webb},
  booktitle={The Routledge handbook of vocabulary studies},
  pages={225--239},
  year={2019},
  publisher={Routledge}
}

@article{xie2024mini,
  title={{Mini-Omni2}: Towards open-source {GPT}-4o with vision, speech and duplex capabilities},
  author={Z. Xie and C. Wu},
  journal={arXiv:2410.11190},
  year={2024}
}

@article{zeng2024glm,
  title={{GLM}-4-voice: Towards intelligent and human-like end-to-end spoken chatbot},
  author={A. Zeng and Z. Du and M. Liu and K. Wang and S. Jiang and L. Zhao and Y. Dong and J. Tang},
  journal={arXiv:2412.02612},
  year={2024}
}

@article{zhang2024internlm,
  title={{InternLM-XComposer2.5-OmniLive}: A comprehensive multimodal system for long-term streaming video and audio interactions},
  author={P. Zhang and X. Dong and Y. Cao and Y. Zang and R. Qian and X. Wei and L. Chen and Y. Li and J. Niu and S. Ding and others},
  journal={arXiv:2412.09596},
  year={2024}
}

@article{yang2025qwen3,
  title={Qwen3 technical report},
  author={A. Yang and A. Li and B. Yang and B. Zhang and B. Hui and B. Zheng and B. Yu and C. Gao and C. Huang and C. Lv and others},
  journal={arXiv:2505.09388},
  year={2025}
}

@article{team2025gemma,
  title={Gemma 3 technical report},
  author={G. Team and A. Kamath and J. Ferret and S. Pathak and N. Vieillard and R. Merhej and S. Perrin and T. Matejovicova and A. Ram{\'e} and M. Rivi{\`e}re and others},
  journal={arXiv:2503.19786},
  year={2025}
}

@inproceedings{qin2014building,
  title={Building a vocabulary self-learning speech recognition system},
  author={L. Qin and A. I. Rudnicky},
  booktitle={Proc. Interspeech},
  address={Singapore},
  year={2014},
  pages={2862--2866}
}

@inproceedings{zhang2020oov,
  title={{OOV} recovery with efficient 2nd pass decoding and open-vocabulary word-level rnnlm rescoring for hybrid asr},
  author={X. Zhang and D. Povey and S. Khudanpur},
  booktitle={Proc. ICASSP},
  address={Barcelona},
  year={2020},
  pages={6334--6338}
}
\endgroup
\end{document}